\documentclass{ws-procs975x65}
\usepackage{amsmath}
\usepackage{chapterbib}

\begin{document}

\title{Holographic Techni-dilaton, or Conformal Higgs\footnote{Talk presented by K.~Haba 
at Nagoya Global COE workshop: ``Strong Coupling Gauge Theories in LHC Era''  in honor of 
Toshihide Maskawa's 70th Birthday and 35th Anniversary of Dynamical Symmetry Breaking in SCGT (SCGT 09),
Dec. 8-11, 2009, Nagoya University, Nagoya, http://www.eken.phys.nagoya-u.ac.jp/scgt09/.
To be published by World Scientific Publishing Co., Singapore (eds. M. Harada, H. Fukaya, M.Tanabashi and K. Yamawaki). }}

\author{$^{a}$Kazumoto Haba, $^{b}$Shinya Matsuzaki, and $^{a}$Koichi Yamawaki}

\address{$^{a}$Department of Physics, Nagoya University,
                    Nagoya, 464-8602, Japan. 
}

\address{$^{b}$Department of Physics,  
                    Pusan National University, Busan 609-735, Korea.
}

\begin{abstract}
We study a holographic model dual to walking/conformal technicolor (W/C TC) deforming   
a hard-wall type of bottom-up setup by including effects from techni-gluon condensation. 
We calculate masses of (techni-) $\rho$ meson, 
$a_1$ meson, and flavor/chiral-singlet scalar meson 
identified with techni-dilaton (TD)/conformal Higgs boson, as well as the $S$ parameter. 
It is shown that gluon contributions and large anomalous dimension tend to 
decrease specifically mass of the TD.  In the typical model with $S \simeq 0.1$, we find
$m_{\rm TD} \simeq 600$ GeV, while $m_\rho,m_{a_1} \simeq 4 {\rm TeV}$.
\end{abstract}

\bodymatter

\section{A holographic technicolor model with techni-gluon condensation} 
The origin of the mass is the most urgent issue of the particle physics today and is 
to be resolved at the LHC experiments. 
In order to account for the origin of mass dynamically without introducing the ad-hoc Higgs boson,
here we consider the Walking/Conformal Technicolor (W/C TC)~\cite{Yamawaki:1985zg,Akiba:1985rr}  having large anomalous dimension $\gamma_m \simeq 1$ due to 
the strong coupling which stays almost non-running (``walking'') over wide energy range near the ``conformal fixed point''. In contrast to a folklore that 
TC is a ``higgsless model'' in analogy with the QCD, there actually exists  ``techni-dilaton' (TD)' ~\cite{Yamawaki:1985zg} as a pseudo Nambu-Goldstone boson of the approximate conformal symmetry, which is relatively light compared with other bound states like techni-$\rho$ and techni-$a_1$, etc..  
Here we shall estimate the mass of TD in the bottom-up holographic W/C TC\cite{}
by newly introducing a flavor-singlet bulk filed corresponding to the techni-gluon condensate. 
Details are given in the forthcoming paper~\cite{HMY2}. 

 Following a bottom-up approach of holographic-dual of QCD~\cite{Erlich:2005qh} 
and that of walking/conformal (W/C) TC~\cite{Hong:2006si,Haba:2008nz},  
we consider a five-dimensional gauge model possessing the $SU(N_f)_L \times SU(N_f)_R$ 
gauge symmetry. 
The model is defined on the five-dimensional anti-de-Sitter space with the 
curvature radius $L$, which is described by the metric $ds^2= g_{MN} dx^M dx^N = 
\left(L/z \right)^2\big(\eta_{\mu\nu}dx^\mu dx^\nu-dz^2\big)$ 
with $\eta_{\mu\nu}={\rm diag}[1, -1, -1,-1]$. 
The fifth direction $z$ is compactified on an interval extended from ultraviolet (UV) 
and infrared (IR) branes, $ \epsilon \leq z \leq z_m  $.  
 In addition to the bulk left- ($L_M$) and right- ($R_M$) gauge fields,  
we introduce a bulk scalar field $\Phi$ which transforms as bifundamental representation under 
the $SU(N_f)_L \times SU(N_f)_R$ gauge symmetry 
so as to deduce the information concerning 
the chiral condensation-operator $\bar{T} T$. According to the holographic dictionary,
the mass-parameter $m_\Phi$ is related to $\gamma_m$, the anomalous dimension of $\bar T T$,  
as $m_\Phi^2=- (3-\gamma_m)(1+ \gamma_m)/L^2$, 
where $\gamma_m \simeq 0$ corresponds to QCD and QCD-like TC and $\gamma_m \simeq 1$ to the W/C TC. 
Here we newly introduce an additional chiral-singlet and 
massless bulk scalar field $\Phi_X$ dual to techni-gluon condensate 
$\langle \alpha G_{\mu\nu}^2 \rangle$, where $\alpha$ is related to the 
TC gauge couping $g_{\rm TC}$ by $\alpha = g_{\rm TC}^2/(4\pi)$. 
We adopt a ``dilaton-like'' coupling 
for interaction terms involving $\Phi_X$ in such a way that 
all the fields couple to $\Phi_X$ in the exponential form like $e^{\Phi_X(z)}$. 
($\Phi_X$ is {\it not} identified with techni-dilaton in this talk.)

Thus the five-dimensional action
takes the form: 
\begin{eqnarray}
S_5 &=&\,\int\,d^4 x\,\int_{\epsilon}^{z_m}\,d\,z~\sqrt{-{\rm det}g_{MN}}\,
\frac{1}{ g_5^2}e^{c g_5^2 \Phi_X(z)} \Big(
-\frac{1}{4}\textrm{Tr}\,\left[{L_{MN}L^{MN}}
+{R_{MN}R^{MN}}\right]
\nonumber \\ 
&& 
\hspace{90pt} 
+\textrm{Tr}\,\left[{D_M\Phi^\dagger D^M\Phi}
-m^2_\Phi \Phi^\dagger \Phi \right]
+ \frac{1}{2} \partial_M 
\Phi_X \partial^M \Phi_X \Big)
\,, 
\label{S5}
\end{eqnarray}
where 
$D_M\Phi=\partial_M \Phi+iL_M\Phi-i\Phi R_M$, 
$g_5$ denotes the gauge coupling in five-dimension, 
and $c$ is the dimensionless coupling constant. 
It turns out that requiring the present model to reproduce 
high-energy behaviors in the underlying theory leads to 
$(L/g_5^2)=N_{\rm TC}/(12\pi^2)$ and $c=-N_{\rm TC}/(192\pi^3)$: 
Our model exactly reproduces the high-energy behaviors for vector and 
axial-vector current correlators up till the terms 
of $1/Q^8$
expected from the OPE. For details, see Ref.3.

We ignore Kaluza-Klein (KK) modes 
of $\Phi_X$ (including the lowest mode) which are identified with massive glueballs 
with mass of order ${\cal{O}}(\Lambda_{\rm TC})$ which is much larger than the electroweak scale, $\Lambda_{\rm TC} \gg F_\pi$, 
in the case of W/C TC with $\gamma_m \simeq 1$. 
The techni-dilaton, a flavor-singlet scalar bound state of
techni-fermion and anti-techni-fermion, will be identified with the lowest KK mode 
in the KK decomposition of $\Phi$, 
$\Phi(x, z) = v(z) + \sigma^{(1)}(x) \sigma_1(z) + \cdots $, but not of $\Phi_X$.

%

Following the holographic recipe, we 
obtain 
formulas for $\langle \alpha G_{\mu \nu}^2 \rangle$, $\langle \bar{T}T \rangle$, 
$M_\rho$, $M_{a_1}$, $M_{\sigma_1} (\equiv M_{\rm TD})$, $F_\pi$, 
and the $S$ parameter, 
where $M_\rho$ and $M_{a_1}$ are the masses of the lowest KK modes 
for the vector and axial-vector mesons identified with the techni-$\rho$ and-$a_1$ mesons. 
Once $\gamma_m$ is specified, 
in the continuum limit $\epsilon \to 0$ 
those quantities involve only three undetermined parameters: $z_m$, and dimensionless quantities
$\xi=\sqrt{2}L \langle \Phi\rangle|_{z=z_m}$ and
$G= \langle e^{c g_5^2 \Phi_X(z)} \rangle|_{z=z_m}-1$
which are related to $\langle \bar{T}T \rangle$ 
and $\langle \alpha G_{\mu \nu}^2 \rangle$, respectively.
Some dimensionless quantities such as $\hat{S}\equiv S/(N_f/2)$ ($S$ parameter per each 
techni-fermion doublet) and $M_{\rho,a_1,{\rm TD}} /F_\pi$ (ratios of meson masses to $F_\pi$)   
are expressed as a function of only $\xi$ and $G$ when $\gamma_m$ fixed. 

\section{Mass of techni-dilaton}
\begin{figure}
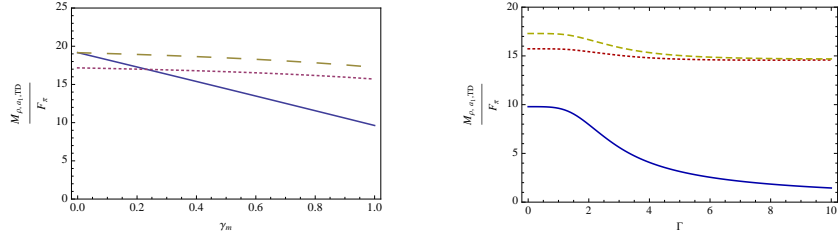

\begin{center}
\includegraphics[width=5.2cm]{mras-ad.eps}
\qquad  \includegraphics[width=5cm]{mras-G.eps}
 \caption{\label{mras:ad}
Plots of $(M_{\rho,a_1, {\rm {TD}}}/F_\pi)$ 
in the case of $\hat S = 0.1$ with 
 $N_{\rm TC} =3$ fixed. The left panel is $\gamma_m$ dependence with
 $\Gamma=1$ and the right panel is $\Gamma$ dependence with $\gamma_m = 1$. 
The dotted, dashed, solid lines 
respectively denote $(M_{\rho}/F_\pi)$, $(M_{a_1}/F_\pi)$,  
and $(M_{\rm TD}/F_\pi)$ in both panels. 
} 
\end{center}
\end{figure}

We first present a generic analysis of the effects  of the gluon condensate 
and the anomalous dimension on $M_{\rho,a_1,{\rm TD}} /F_\pi$. 
The gluon contribution may be 
evaluated through a quantity 
$\Gamma = \Gamma(\gamma_m,\xi,G) \equiv 
\left(\left(\frac{1}{\pi}\langle \alpha G_{\mu \nu}^2 \rangle/F_\pi^4\right)/
\left(\frac{1}{\pi}\langle \alpha G_{\mu \nu}^2 \rangle/f_\pi^4\right)_{\rm QCD}\right)^{1/4}
$. 
Fig.~\ref{mras:ad} shows plots of $M_{\rho,a_1,{\rm TD}} /F_\pi$
as a function of $\gamma_m$ and $\Gamma$ and indicates that 
$M_{\rm TD}/F_\pi$ rapidly decreases  
as  $\gamma_m$ and/or $\Gamma$  become larger in contrast to $M_{\rho,a_1}/F_\pi$. 

We now consider  
a typical model of W/C TC,  
based on the Caswell-Banks-Zaks infrared fixed point 
in the large $N_f$ QCD where we use 
the estimate $N_f \simeq 4 N_{\rm TC}=12 (N_{\rm TC} = 3)$ 
from the two-loop beta function 
and ladder Schwinger-Dyson equation. 
In the W/C TC dynamics 
the techni-gluon condensate   
is not on order of the intrinsic scale $\Lambda_{\rm TC}$ but of the mass of techni-fermions $m (\ll \Lambda_{\rm TC})$ through the 
conformal anomaly associated with the mass generation, $\partial^\mu D_\mu =\theta^\mu_\mu=4\theta^0_0 = \beta(\alpha)/(4\alpha^2)\cdot\langle
\alpha G_{\mu\nu}^2\rangle$ and $\langle \theta_0^0\rangle = -\frac{N_f N_{\rm TC}}{\pi^4}m^4$,
where the nonperturbative beta function behaves as $\beta(\alpha) \rightarrow 0$ when $\Lambda_{\rm TC}/m \rightarrow \infty$ (conformal fixed point).
This implies that  the gluon condensate $\langle
\alpha G_{\mu\nu}^2\rangle$ (or $\Gamma$) becomes large in that limit and hence $m_{\rm TD} \rightarrow 0$ as is seen in Fig.~\ref{mras:ad}.
Indeed $\Lambda_{\rm TC}$ is identified with the Extended TC (ETC)  scale $\Lambda_{\rm ETC}$:
$\Lambda_{\rm TC}=\Lambda_{\rm ETC} \sim (10^4-10^5) F_\pi$ where  $F_\pi(={\cal O}(m))$ is given by $F_\pi=246/\sqrt{N_f/2}$ GeV. Using  the phenomenological
input $S=0.1$, we have  $\Gamma\simeq 7$. 
%
Then we find $M_{\rm TD}
\simeq 550$--$680$ GeV
in contrast to $M_\rho \simeq M_{a_1} \simeq$ $3.8$--$3.9$ TeV. 

\section*{Acknowledgments}

This work was supported in part by 
the JSPS Grant-in-Aid for Scientific Research: (B) 18340059, 
the Global COE Program
``Quest for Fundamental Principles in the Universe''
 and the Daiko Foundation. S.M. was supported by the Korean Research Foundation Grant
  funded by the Korean Government
 (KRF-2008-341-C00008).

\bibliographystyle{ws-procs975x65}
\bibliography{ws-pro-sample}

\end{document}